\documentstyle[12pt,prd,aps]{revtex}
\def\be{\begin{equation}}
\def\en{\end{equation}}
\def\bea{\begin{eqnarray}}
\def\ena{\end{eqnarray}}
\newcommand{\e}{\mbox{e}}
\newcommand{\J}{\mbox{J}}
\newcommand{\sgn}{\mbox{sgn}}
\begin{document}
\title{HIGH ENERGY PARTICLES FROM \\ MONOPOLES CONNECTED BY
STRINGS} \author{Veniamin BEREZINSKY}\address{INFN, Laboratori Nazionali
del Gran Sasso,\protect\\ I-67010 Assergi (AQ), Italy}
\author{Xavier MARTIN and
Alexander VILENKIN} \address{Institute of Cosmology, Department of
Physics and Astronomy,\protect\\ Tufts University, Medford, MA
02155, USA} \maketitle

\begin{abstract}
{\sl Monopole-antimonopole pairs connected by strings and monopole-string
networks with $N>2$ strings attached to each monopole can be formed at
phase transitions in the early universe.  In such hybrid defects,
monopoles accelerate under the string tension and can reach
ultrarelativistic Lorentz factors, $\gamma\gg 1$.  We study the
radiation of gauge quanta by accelerating monopoles.  For monopoles
with a chromomagnetic charge, we also discuss the high-energy hadron
production through emission of virtual gluons and their subsequent
fragmentation into hadrons. The relevant
parameter for gauge boson radiation is $M/a$, where $M$ is the boson
mass and $a$ is the proper acceleration of the monopole.  For $M\ll
a$, the gauge bosons can be considered as massless and the typical
energy of the emitted quanta is $E\sim\gamma a$.  In the opposite
limit, $M\gg a$, the radiation power is exponentially suppressed and
gauge quanta are emitted with a typical energy $E\sim\gamma M$ in a
narrow range $\Delta E/E\sim (a/M)^{1/2}$.  Cosmological
monopole-string networks can produce photons and hadrons of extremely
high energies.  For a wide range of parameters these energies can be
much greater than the Planck scale.}
\end{abstract}

\section{Introduction}
Monopoles connected by strings can be formed in a sequence of symmetry
breaking phase transitions in the early universe \cite{kibble,shelvil}. 
A typical sequence of this sort is \be
G\rightarrow H\times U(1)\rightarrow H\times Z_N .\label{symbreak} \en
For a semi-simple group $G$, the first of these phase transitions
gives rise to monopoles, and at the second phase transition each
monopole gets attached to $N$ strings. For $N\geq 3$, a single
infinite network is formed which permeates the entire universe
\cite{aev}. The magnetic fluxes of monopoles in the network are
channelled into the strings that connect them. But the monopoles
typically have additional, unconfined magnetic charges. The gauge
fields associated with these charges may include electromagnetic or
gluon fields, but may also correspond to broken gauge symmetries and
have non-zero masses.

The cosmological evolution of monopole-string networks has been
discussed in Ref. \cite{vv}, where it was argued that the networks
evolve in a self-similar manner, with a characteristic scale growing
proportionally to cosmic time $t$. The typical energy of the monopoles
also grows like $t$, and the monopoles become highly relativistic.
Radiation of gauge quanta by accelerating monopoles is the main energy
loss mechanism of the networks. Relativistic particles emitted by the
monopoles (or the decay products of these particles) contribute to the
spectrum of high-energy cosmic rays.

For the symmetry breaking sequence \be G\rightarrow H\times
U(1)\rightarrow H,\label{symbreak1} \en which can be regarded as a
special case of (\ref{symbreak}) with $N=1$, the cosmological
evolution is quite different. In this case, each monopole is attached
to a single string, and the second phase transition results in
monopole-antimonopole (M\={M}) pairs connected by strings. If both of
the phase transitions in (\ref{symbreak1}) occur during the
radiation era, then the average monopole separation is always smaller
than the Hubble radius, and when M\={M} pairs get connected by strings
and begin oscillating, they typically dissipate the bulk of their
energy to friction in less than a Hubble time \cite{shelvil,hkr}.

A more interesting possibility arises in the context of inflationary
scenario, when monopoles are formed during inflation but are not
completely inflated away. Strings can either be formed later during
inflation, or in the post-inflationary epoch. In this case, the
strings connecting M\={M} pairs can be very long. The correlation
length of strings, $\xi$, can initially be much smaller than the
average monopole separation, $d$; then the strings connecting
monopoles have Brownian shapes. But in the course of the evolution,
$\xi$ grows faster than $d$, due to small loop production, and to the
damping force acting on the strings. Eventually, $\xi$ becomes
comparable to the monopole separation, and we are left with $M\bar{M}$
pairs connected by more or less straight strings. At later times, the
pairs oscillate and gradually lose their energy by gravitational
radiation and by radiation of light gauge bosons (if the monopoles
have unconfined magnetic charges). When the energy of a string
connecting a pair is dissipated, the monopoles and antimonopoles
annihilate into relativistic particles.

Although monopole-antimonopole pairs connected by strings do not
typically survive until the present time, they can produce a
characteristic feature in the spectrum of the stochastic gravitational
wave background \cite{vilrev}. The magnitude of this feature depends
in particular on the intensity of the gauge boson radiation, which
determines the lifetime of the pairs.

The production of high energy particles by topological defects 
is interesting as a mechanism of emission
of ultrahigh energy particles, in particular photons, nucleons and
neutrinos. The energies of such particles can be much higher than have
been observed until now. Particle production by cosmic strings
\cite{W+} and by annihilating monopole-antimonopole pairs \cite{H+}
has been previously discussed in the literatute.  Here, we shall
calculate the radiation power
and the spectrum of gauge bosons, both massless and massive, emitted
by an oscillating monopole--antimonopole pair. If the magnetic charge
of the monopoles is of electromagnetic nature, then the radiation can
be treated in the same way as that of an accelerating charge, and our
problem reduces to a problem in classical electrodynamics. In the case
of a chromomagnetic charge, the picture is more complicated. The
monopoles emit heavy virtual gluons which originate a parton
cascade. The hadrons are produced in the usual way beyond the
confinement radius. The decay of pions results in high energy neutrino
radiation. The heavy gauge bosons, such as $W$ and $Z$, can also be
produced for some range of parameters of the M\=M pair and the string.

The problem studied here is in fact a general theoretical problem: the
radiation of heavy particles by an accelerating monopole. Our results
can be reformulated in terms of the basic parameters of a shock problem,
the proper acceleration of the monopole.

The dynamics of M\=M pairs connected by strings is rather complicated
and has not been studied in any detail in the litterature. We shall
therefore concentrate on the simplest case of an oscillating pair
connected by a straight string, for which the equations of motion can
be solved exactly. In addition, we shall consider the case of a
harmonic oscillator motion of the monopoles which has some of the
features expected in more general monopole-string
configurations. Taken together, the analyses of these two cases will
allow us to draw a reasonably complete picture of the radiation
spectrum in the general case.

After reviewing the general formalism for gauge boson radiation in the
next Section, we calculate the radiation spectrum of massless gauge
bosons in Section \ref{sec3}. Radiation of massive gauge bosons is
analysed in Section \ref{sec4}. The spectrum of hadrons resulting from
radiation of gluons and their subsequent fragmentation are discussed in
section \ref{sec5}. Finally, Section \ref{sec6} contains some
concluding remarks.

\section{\bf General formalism}

The characteristic monopole radius $\delta_m$ and string thickness
$\delta_s$ are determined primarily by the corresponding symmetry
breaking scales, $\eta_m$ and $\eta_s$. Typically, $\delta_m \sim
\eta_m^{-1}$ and $\delta_s \sim \eta_s^{-1}$. We shall assume that 
the two symmetery breaking scales
are well separated, $\eta_s \ll \eta_m$, and thus the monopole
radius is much smaller than the string thickness, $\delta_m \ll
\delta_s$.

Assuming also that the string length is much greater than its
thickness, we can treat monopoles as point particles and strings as
infinitely thin lines. The monopole equations of motion can then be
written as \cite{carter} \be 
\frac{d^2 x^\nu}{ds^2}= \frac{\mu}{m} \lambda^\nu ,\label{moneq} \en
where $\mu$ is the mass per unit length (and tension) of the
Goto-Nambu string, $m$ is the monopole mass, $\tau$ is the proper time
of the monopole, and $\lambda^\nu$ is a unit vector orthogonal to the
monopole wordline and oriented into the string worldsheet. It follows
immediately from Eq. (\ref{moneq}) that
\be a=\frac{\mu}{m},\label{mrate} \en
is the proper acceleration of the monopoles. Eq. (\ref{moneq}) can
also be rewritten in a $3$-dimensional form: \be
\gamma^3 \ddot{\bf x} (t) =a\gamma^{(s)} {\bf n}. \label{moneq2} \en
where, $\gamma =(1-\dot{\bf x}^2 )^{-1/2}$ is the Lorentz factor of
the monopole, $\gamma^{(s)}$ is the Lorentz factor of the string at
the location of the monopole, and ${\bf n}$ is a unit vector
pointing from the monopole in the direction of the string.
Eqs (\ref{moneq}) and (\ref{moneq2}) have a simple physical meaning: in
the instantaneous rest frame of the monopole, the magnitude of its
acceleration is given by Eq. (\ref{mrate}) and the direction is given
by the direction of the string. The motion of the string is described
by the usual Goto-Nambu equations. A simple solution of
Eq. (\ref{moneq2}) describing an oscillating M\=M pair connected by a
straight string is \cite{smgr} \begin{mathletters} \label{strsol} \bea
x (t) & = & \pm \frac{\sgn (t)}{a} \left[ \gamma_0 -\sqrt{1+ (\gamma_0 \
v_0 -a|t|)^2} \right] , \\
y (t) & = & z (t) =0 , \ena \end{mathletters}
where $v_0$ and $\gamma_0 =(1-v_0^2 )^{-1/2}$ are respectively the
maximal velocity and the maximal Lorentz factors of the monopoles,
reached at $t=0$. The upper and lower signs correspond to M and \=M
respectively.

The solution (\ref{strsol}) is valid for $|t|\leq \gamma_0 v_0 /a$. At
$t=-\gamma_0 v_0 /a$, the monopoles are at rest, with the string
having its maximum length, $L=2(\gamma_0 -1)/a$. At $t=+\gamma_0 v_0
/a$, the monopoles come to rest again, with their positions
interchanged. Eq. (\ref{strsol}) describes only half a period,
$T=4\gamma_0 v_0 /a$, the other half period being obtained by
exchanging the positions of the monopole and antimonopole. A peculiar
feature of this solution (\ref{strsol}) is that the monopole
accelerations abruptly change direction when the monopoles meet and
pass one another. Between these sign changes the monopoles move at
constant proper acceleration $a$. The monopole and antimonopole meet at
$t=0$ and could be expected to annihilate. However, this solution is
considered as an approximation for an almost straight configuration
where the monopole and antimonopole would merely come close to each
other and would not collide. Besides, as we already mentioned, the
monopole radii are much smaller than the string thickness, thus the
monopoles are not likely to collide, even for a straight string.

To find the radiation power of gauge bosons by accelerating monopoles,
we first consider radiation by ordinary gauge charges. The gauge field
$A^\mu (x)$ produced by the $4$-current $j^\mu (x)$ is determined by
the equations \be
{F^{\nu\sigma}}_{,\nu} +M^2 A^\sigma = j^\sigma ,\label{poteq} \en
where $F_{\nu\sigma} =\partial_\nu A_\sigma -\partial_\sigma A_\nu$ is
the field strength and $M$ is the gauge boson mass. The current
corresponding to a pair of equal and opposite point charges is \be
j^\nu (x) = q\dot{x}^\nu_+ \delta^{(3)} ({\bf x}-{\bf x}_+ (t))-
 q\dot{x}^\nu_- \delta^{(3)} ({\bf x}-{\bf x}_- (t)), \label{jdef} \en
where ${\bf x}_+ (t)$ and ${\bf x}_- (t)$ are the trajectories of the charges and $\dot{x}^\nu_\pm = dx^\nu_\pm /dt$. In the case of a periodic motion, the radiation power can be found from the following equations \be
P=\sum_n P_n = \sum_n \int d\Omega \frac{dP_n}{d\Omega}, \label{Pdef} \en \be
\frac{dP_n}{d\Omega} = -\frac{k\omega_n}{8\pi^2} j^*_\mu (\omega_n ,{\bf k})
j^\mu (\omega_n ,{\bf k}) .\label{bospec} \en
Here, $dP_n /d\Omega$ is the radiation power at frequency $\omega_n =2
\pi n/T$ per unit solid angle in the direction of $\bf k$, $|{\bf k}|=
(\omega_n^2 -M^2)^{1/2}$, $T$ is the period of the oscillation, and \be
j^\mu (\omega_n ,{\bf k})=\frac{1}{T} \int_0^T dt \exp (i\omega
_n t)\int d^3 x \exp (-i{\bf k}\cdot{\bf x}) j^\mu ({\bf x},t)
\label{jfour} \en 
is the Fourier transform of the current.  Eq. (\ref{bospec}) is
derived in the Appendix \ref{bospow}. The derivation is similar to
that of the gravitational radiation power in Weinberg's book
\cite{wein}.

Because of electric-magnetic duality, the same equations apply to
radiation by magnetic monopoles. Hence, the radiation power from an
oscillating M\=M pair can be found from Eqs. (\ref{jdef})-(\ref{jfour})
with $q$ interpreted as the magnetic charge.

For a one dimensional motion, the current has only two non-zero
components: $j^0$ and $j^1$. It is also conserved, $\nabla_\nu j^\nu
=0$, which in Fourier space can be written as \be \omega_n j^0 = k_x
j^1 ,\en where $k_x \equiv k^1$. Combining this with
Eq. (\ref{bospec}), we have \be 
\frac{dP_n}{d\Omega} =\frac{\omega_n k}{8\pi^2} (1-\frac{k_x^2}{\omega
_n^2}) |j_1 (\omega_n ,{\bf k})|^2 .\label{powspec2} \en
The motion of the M\=M pair in Eq. (\ref{strsol}) has the properties 
\begin{mathletters} \bea x_+ (t) & = & -x_- (t)=-x_+ (-t) \\ 
x_+ (t) & = & -x_+ (t+T/2). \ena \end{mathletters}
In such a case, it is easily seen that the system radiates only in odd
modes, and that for $n$ odd, we have \be
j_1 (\omega_n ,{\bf k}) =\frac{8q}{T}\int_0^{T/4} \dot{x} (t) dt
\cos (\omega_n t) \cos [k_x x_+ (t)]. \label{powsym} \en

Before we proceed with the calculation of the radiation spectrum, we
shall briefly comment on the limits of applicability of the classical
treatment of the radiation process that we adopt in this paper.  The
problem of radiation by a monopole pulled by a contracting string is
equivalent to that of a monopole moving in a longitudinal magnetic
field, or of an electric charge moving in a longitudinal electric
field.  Such processes in strong
fields are characterized \cite{Ritus},\cite{Berest} by the invariant
$\chi$ given by 
\be
\chi=(q/m^3)\sqrt{(F_{\mu\nu}p_{\nu})^2} =f/m^2, \label{chi} \en
where $q$, $m$ and $p_{\nu}$ are respectively the charge, mass and 
momentum of the radiating particle and $f$ is the force acting
on the particle in its instantaneous rest frame.  In the case of a
monopole connected to a string the corresponding parameter is
$\chi=\mu /m^2=a/m$.
The case $\chi \ll 1$ corresponds to the quasiclassical regime
\cite{Ritus}. Hence, our classical treatment of radiation is justified
provided that $\mu \ll m^2$.

\section{Radiation of massless gauge bosons} \label{sec3}
We begin by calculating the radiation of massless gauge bosons by an
oscillating M\=M pair described by Eq. (\ref{strsol}). In this case,
Eqs. (\ref{powspec2}), (\ref{powsym}) for the radiation power can be
transformed to \bea
P_n & = & \frac{2q^2 a^2}{\pi\gamma_0^2 v_0^2} \int_0^1 (1-u^2 ) [S_n
(u)]^2 du,\label{jinta}\\
S_n (u) & = & \int_0^{\frac{n\pi}{2v_0}(1-\frac{1}{\gamma_0})} \cos (u\xi )
\sin \left( \sqrt{\left( \frac{n\pi}{2v_0}-\xi \right) ^2 -\frac{n^2
\pi^2}{4\gamma_0^2 v_0^2}}\right) d\xi .\label{jintb} \ena
where we have introduced new integration variables $u=k_x /\omega_n$
and $\xi =n\pi a x_+ (t)/2\gamma_0 v_0$. These integrals cannot be
evaluated analytically. However, the high-frequency asymptotic
behaviour can be obtained as an expansion in powers of $1/n$. This can
be done simply by repeatedly integrating by parts in Eq. (\ref{jintb})
\cite{smgr}.

The leading term of the expansion is \be
S_n (u) \approx \frac{2v_0}{n\pi\gamma_0^2} \frac{1+3u^2 v_0^2}{(1-u^2
v_0^2 )^3} \en
and the corresponding power spectrum is \be
P_n \approx A\left( \frac{qa\gamma_0}{n}\right) ^2 ,\en
where \be A=\frac{8}{\pi^3} \int_0^1 (1-x^2) \frac{[\gamma_0^2+3x^2
(\gamma_0^2-1)]^2}{[\gamma_0^2-x^2 (\gamma_0^2-1)]^6}dx.\en
For ultrarelativistic monopoles, $\gamma_0 \gg 1$, this can be 
approximated as \be
P_n \approx \frac{16}{5\pi^3}\left( \frac{qa\gamma_0}{n}\right)^2 .
\label{asympem} \en
The $1/n$ expansion is valid as long as the first neglected term is
small compared to the leading term (\ref{asympem}). A somewhat lengthy
but staightforward calculation gives the condition $n\gg
\gamma_0^2$. At lower frequencies, the $n$-dependence is rather
complicated and analytic approximations are difficult to obtain.

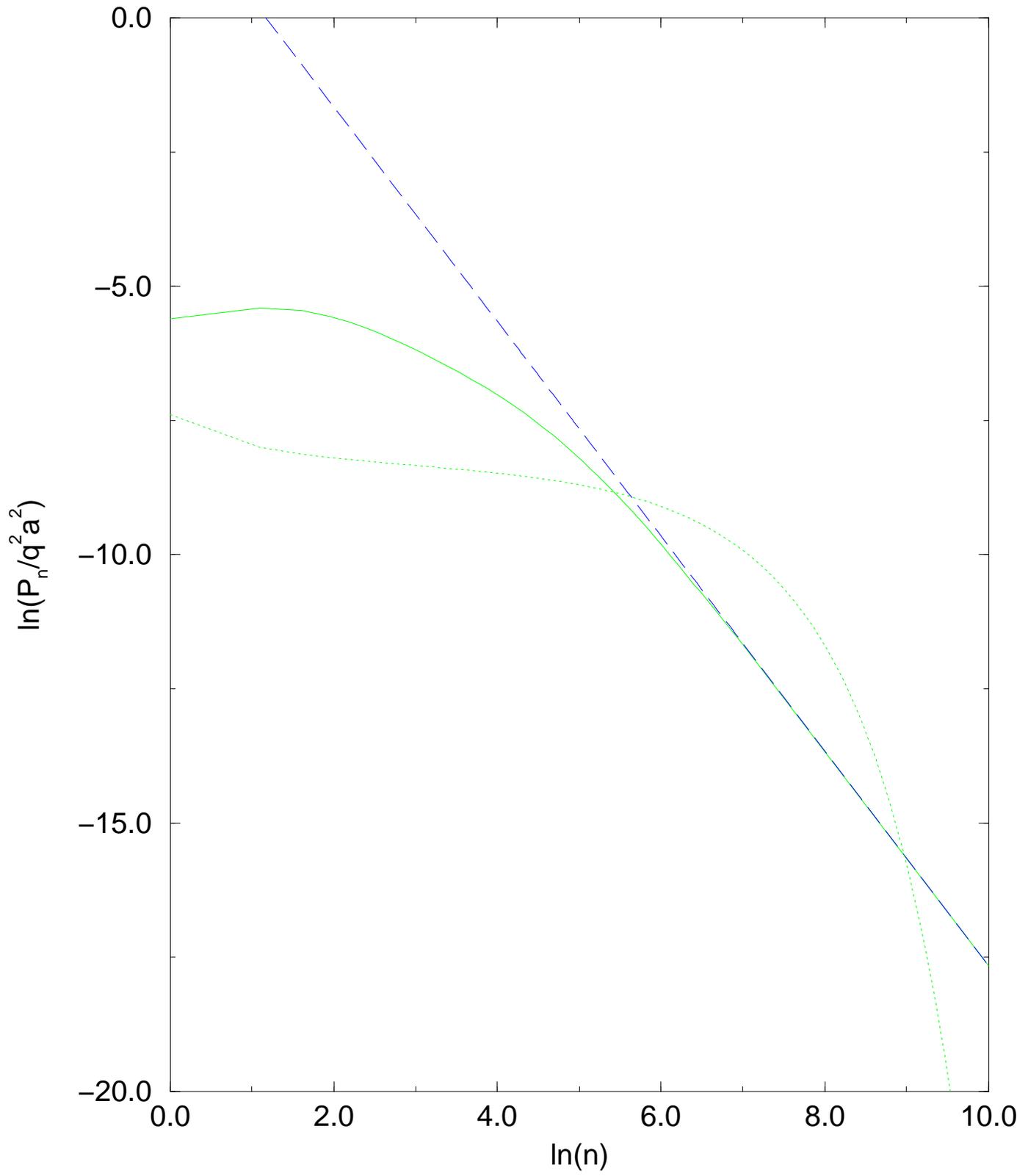
\begin{figure}
\caption{A log-log plot of the gauge quanta radiation spectrum $P_n
/(qa)^2$ in the case $\gamma_0 =10$ (solid line) with its high
frequency approximation (\protect\ref{asympem}) (dashed line). The
corresponding spectrum for the harmonic solution has been added
(dotted line) for comparison.}
\label{gamem10}
\end{figure}

We computed the radiation spectrum numerically by integrating
Eqs. (\ref{jinta}), (\ref{jintb}) for various values of $\gamma_0$. A
generic example of this spectrum, obtained for $\gamma_0=10$, is
plotted in Figure \ref{gamem10}. It exhibits the expected behaviour
(\ref{asympem}) at high frequencies and is approximately flat at lower
frequencies. The transition happens around $n\sim \gamma_0^2$ which is
the characteristic frequency $\overline{\omega} \sim \gamma_0^2
\omega_1$, at which most of the power is radiated. We note that in the
rest frame of the monopole it is \be
\omega_{rest} \sim \overline{\omega}/\gamma_0 \sim \gamma_0 \omega_1
\sim a.\en

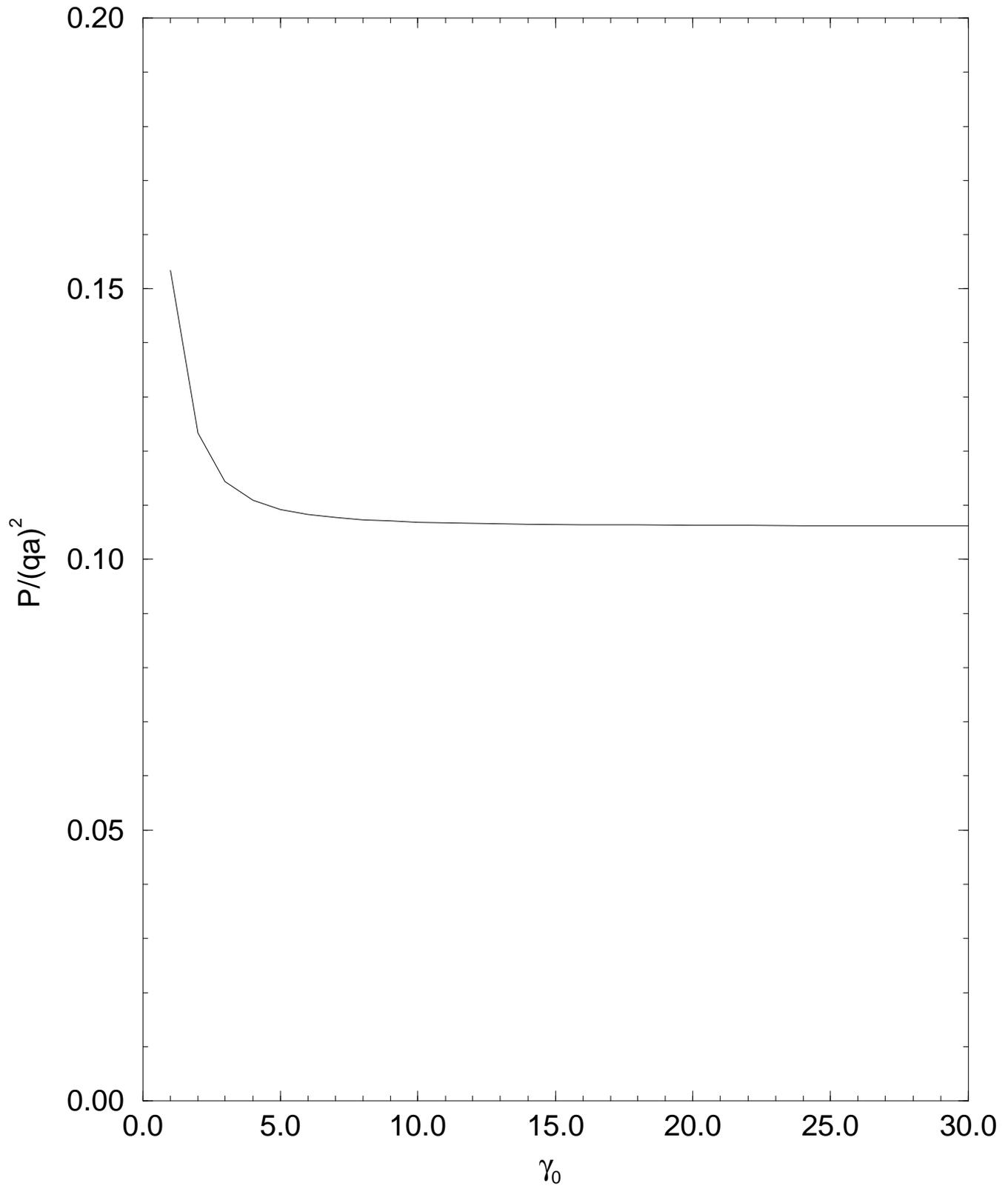
\begin{figure}
\caption{Total power of massless gauge quanta emission
$P/(qa)^2$ as a function of $\gamma_0$. It quickly tends to a limit of
$1/3\pi$.}
\label{totempow}
\end{figure}

The total power radiated by the system can be evaluated directly using
the standard formulas of electrodynamics, \be
P=\frac{(qa)^2}{4\pi} \int_0^1 du(1-u^2 )\int_0^1 d\xi ([\gamma_0^2 (u
-v_0\xi)^2 +1-u^2]^{-3/2}+[\gamma_0^2(u+v_0\xi)^2+1-u^2]^{-3/2})^2.\en
When the monopole and antimonopole reach ultrarelativistic speeds,
their radiation becomes focused in their respective forward
directions, which are opposite, and thus interference between the
monopole and antimonopole fields is negligible. This means that the
total power radiated by the system is close to twice the power
radiated by the monopole alone. The radiation intensity of a single
charge $q$ moving with proper acceleration $a$ is \cite{ll} $P=q^2
a^2/6\pi$.  Thus, for high velocities of the monopole and antimonopole
$\gamma_0 \gg 1$, the total power tends to \be
P\approx \frac{1}{3\pi}(qa)^2 . \label{totemasymp} \en
The numerically calculated power has been plotted as a function of
$\gamma_0$ in Figure \ref{totempow} and indeed quickly tends to this
asymptotic limit.

The monopole trajectory (\ref{strsol}) that we studied so far is
rather special. As we already mentioned, the monopole accelerations
change discontinuously as the monopoles pass one another. It is this
feature that is responsible for the power-law (rather than
exponential) falloff of $P_n$ at large $n$. To see how the character
of the spectrum is changed in a more generic case when the
acceleration varies smoothly, we considered a simple
harmonic-oscillator-type motion of the monopoles, \be
x_+ (t) =x_0 \sin \Omega t =-x_- (t) .\label{harmsol} \en
The period of this motion is $T=2\pi /\Omega$, the maximum velocity
reached by the monopoles is $v_0 =\Omega x_0$, and the maximum Lorentz
factor is $\gamma_0 =(1-v_0^2)^{-1/2}$. The parameters $x_0$ and
$\Omega$ should of course satisfy $v_0 =\Omega x_0 <1$, but we can
shoose $v_0$ to be arbitrarily close to $1$. The proper acceleration
of the monopoles in Eq. (\ref{harmsol}) is \be
a(t)=-\frac{\Omega^2x_0\sin \Omega t}{[1-\Omega^2x_0^2 \cos^2 \Omega t
]^{3/2}}.\label{accprop} \en

Its maximum  value determined by condition $\dot{a}(t_m)=0$ is reached at 
$\sin \Omega t_m=\pm1/\sqrt2\gamma_0v_0$, if $v_0 > 1/\sqrt3$. It is
given by
\be a_m =\frac{2}{3\sqrt3} \gamma_0^2 \Omega .\en

For ultrarelativistic monopoles, $\gamma \gg 1$, the acceleration
remains of this order of magnitude, $a\sim \gamma_0^2\Omega$, only for 
a brief interval of time, \be
\Delta t \sim \left( \frac{a_m}{|{\ddot a}_m|}\right)^{1/2} 
\sim (\gamma_0 \Omega )^{-1}. \label{deltat} \en
Since $ P \propto a_m^2$ this part of the oscillation period gives the dominant
contribution to the radiation power in the massless case.

The power radiated in massless gauge bosons by the harmonic solution
(\ref{harmsol}) can be computed from the standard formulas
(\ref{powspec2}) and (\ref{powsym}). It gives \be
P_n =\frac{2}{\pi} (n\Omega q)^2 \int_0^1 du (\frac{1}{u^2}-1) \J_n^2
(nuv_0),\label{harmpow} \en 
where $\J_n$ is the $n$-th Bessel function.

For large frequency modes $n\gg 1$ and a parameter $x<1$, the Bessel
function can be expanded using the formula \be
\J_n^2 (nx)\simeq \frac{1}{2n\pi} \frac{\e^{-2n(1-x^2)^{3/2}/3}}{
\sqrt{1-x^2}} \label{bessexp} \en
to get the simplified expression for the power \be
P_n \approx n\left( \frac{\Omega q}{\pi}\right) ^2 \int_0^1
du (\frac{1}{u^2}-1)\frac{\e^{-2n(1-u^2v_0^2)^{3/2}/3}}{\sqrt{1-u^2v_0
^2}}.\label{app1} \en
At high frequencies, the radiation is focused in the monopoles forward
direction, which corresponds to $u\sim 1$. This translates in
(\ref{app1}) to the steep fall off of the exponential.

For not too high frequency modes $1\ll n\lesssim \gamma_0^2$, the
expression of the Fourier transform of the current can be somewhat
simplified \bea
j_0 (n\Omega ,{\bf k}) & = & \frac{8q}{T} \int_0^{T/4} dt
\sin (n\Omega t) \sin [ nu (1-1/\gamma_0^2 )\sin (\Omega t)], \nonumber \\
& \approx &  \frac{4q}{\pi} \int_0^{\pi/2} dv \sin (nv) \sin [ nu
\sin v], \ena
where $u=k_x/\Omega$. This corresponds basically to taking $v_0=1$ in
$j_0$. Thus, in this frequency range, the approximation (\ref{app1})
takes the form \be
P_n \approx n \left( \frac{\Omega q}{\pi} \right) ^2 \int_0^1 \frac{du
}{u^2} \sqrt{1-u^2} \exp \left( -\frac{2n}{3}(1-u^2)^{3/2} \right) .
\label{lnapp1} \en
A simple change of variable $\xi =2n(1-u^2)^{3/2}/3$ enables us to
evaluate the integral: \bea
P_n & \approx & \frac{(\Omega q)^2}{2\pi^2} \int_0^{2n/3} \frac{d\xi \e
^{-\xi}}{[1-(3\xi /2n)^{2/3}]^{3/2}} \nonumber \\
& \approx & \frac{(\Omega q)^2}{2\pi^2} \int_0^\infty d\xi \e^{-\xi}
= \frac{(\Omega q)^2}{2\pi^2}.\label{lnharmpow} \ena
So the spectrum is expected to be flat in this frequency range.

For very high frequency modes $n\gg \gamma_0^3$, we can expand the
argument of the exponential in (\ref{app1}) around $u=1$, and simply
take all the other slower varying terms as being constant. This gives
\bea P_n & \approx & \frac{2}{\pi^2} (\Omega q)^2 n\e^{-2n/3\gamma_0^3}
\int_0^{+\infty} vdv \e^{-2nv_0^2 v/\gamma_0} \nonumber \\
& \approx & \frac{\gamma_0^3\Omega^2q^2}{2\pi^2 v_0^4}
\frac{\e^{-2n/3\gamma_0 ^3}}{n}. \label{harmasympem} \ena

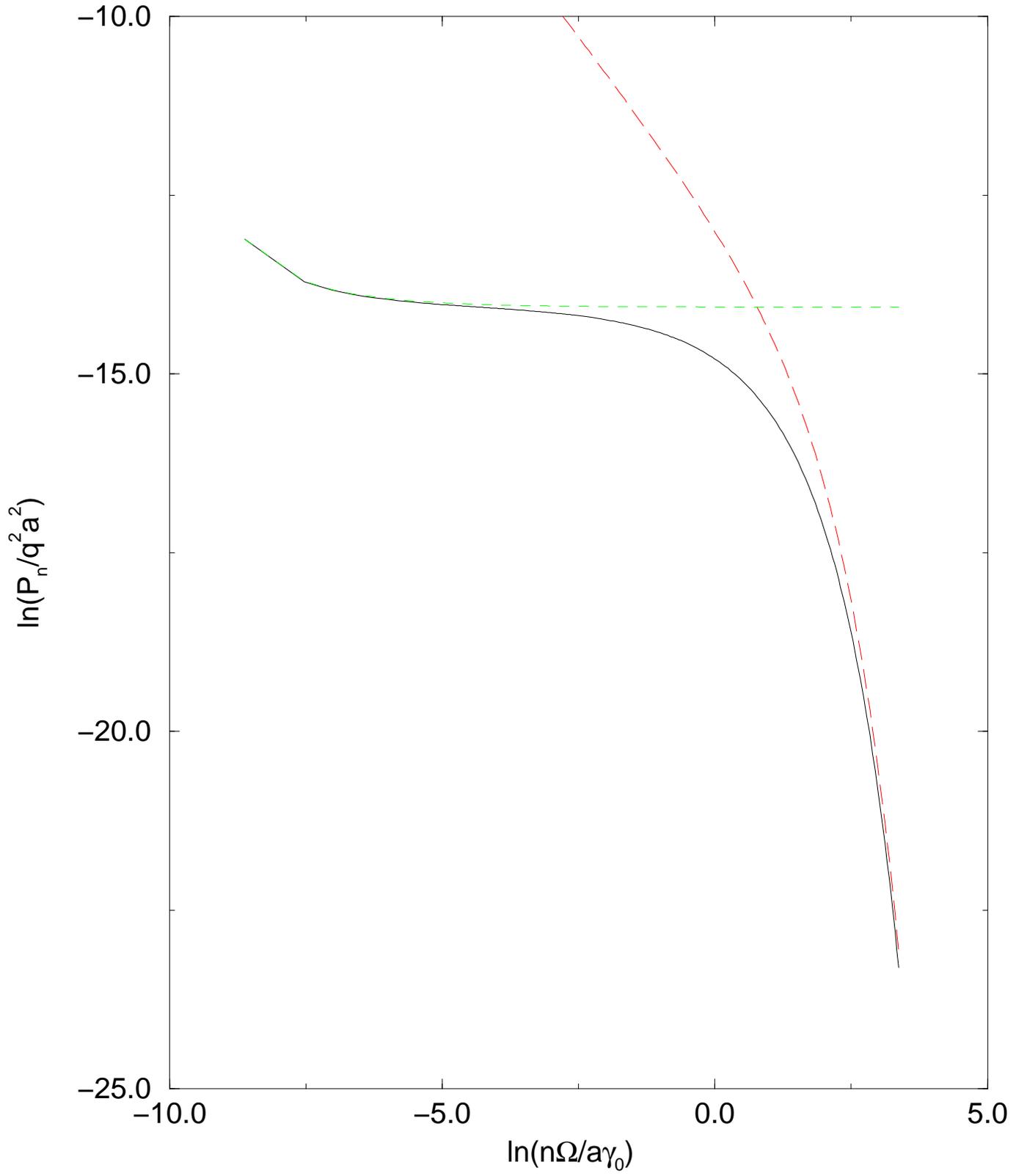
\begin{figure}
\caption{A log-log plot of the gauge quanta radiation spectrum $P_n
/(qa)^2$ from the harmonic solution (\protect\ref{harmsol}) in the
case $\gamma_0 =25$ (solid line) with its high frequency
(\protect\ref{harmasympem}) (long dashed line) and low frequency
(\protect\ref{lnharmpow}) (dashed line) approximations.}
\label{harmpow25}
\end{figure}

The radiation spectrum (\ref{harmpow}) was computed numerically for
various values of $\gamma_0$. A generic example of this spectrum
obtained for $\gamma_0 =25$ is given in Figure \ref{harmpow25}. As
found in (\ref{lnharmpow}), the low frequency behaviour is flat, up to
frequencies ${\bar\omega} \sim a_m \gamma_0$. The expected high
frequency behaviour (\ref{harmasympem}) then sets in at higher
frequencies. Compared to the spectrum obtained for the exact solution
(\ref{strsol}), this spectrum is very similar: flat at lower
frequencies then decaying fast. But, for the harmonic solution, the
flat part has a smaller value, extends farther in frequencies and the
fall off at higher frequencies is much faster.

The total power radiated by the two monopoles is easy to evaluate when
they are ultrarelativistic $\gamma_0\gg 1$. In this limit, as
explained previously, interference between the monopole and
antimonopole fields is negligible and the total power is just the sum
of the power emitted by the monopole and antimonopole separately. The
radiation intensity for a single charge moving with a proper
acceleration (\ref{accprop}) is \cite{ll} $P=9 q^2 a_m^2/16
\gamma_0$. Thus, for ultrarelativistic monopoles $\gamma_0\gg1$, the
total power tends to \be 
P\approx \frac{9}{8\gamma_0} q^2 a_m^2 =\frac{1}{6} \gamma_0^3 \Omega
^2 q^2.\label{harmemtot} \en 
Contrary to the previous case, Eq. (\ref{totemasymp}), this has an
extra factor $\gamma_0$ in the denominator. It comes from the fact
that for ultrarelativistic monopoles, the main contribution to the
total power comes from the interval of time $\Delta t$ in
(\ref{deltat}) when their acceleration is near its maximum $a_m$ so
that $P\sim q^2 a_m^2 \Delta t /T \sim q^2 a_m^2 /\gamma_0$. This
total power is also consistent with the picture of a spectrum
consisting of a low frequency flat behaviour $P_n \sim A\gamma_0^{-4}$
up to $n\sim \gamma_0^3$ as shown in (\ref{lnharmpow}), and then a
negligible exponentially decreasing high frequency behaviour which can
be ignored. This gives a useful approximation for the spectrum of
massless gauge bosons, \begin{mathletters} \label{specap} \bea
P(E) & = & \frac{(\Omega q)^2}{2\pi^2} \equiv P_0 \ \  \mbox{ for }
\ \ E<\frac{\pi ^2}{3} \gamma_0^3\Omega \equiv E_0 ,\\ 
P(E) & = & 0\ \ \ \ \ \ \ \ \ \ \ \ \ \ \ \ \mbox{ for } \ \ E>E_0 ,
\ena \end{mathletters}
which has the same total power as the exact spectrum and will be a
good approximation for large $\gamma$-factors.

It is instructive to compare the results we obtained in this Section
with the case of synchrotron radiation by a charge moving at a
constant speed in a circular orbit \cite{ll}.  The power spectrum in
this case exhibits a peak around ${\bar\omega}\sim\gamma^3\Omega$,
where $\gamma$ is the Lorentz factor and $\Omega$ is the angular
frequency of the rotating charge.  The proper acceleration of the
charge is $a\approx\gamma^2\Omega$, so once again the characteristic
frequency is ${\bar\omega}\sim\gamma a$.   

\section{Radiation of massive gauge bosons} \label{sec4}

We will now consider the radiation of massive gauge bosons of mass
$M$. For simplicity, we will consider only the harmonic oscillation
solution (\ref{harmsol}) in this section.

The power radiated in massive gauge bosons is still given by the
general formula (\ref{powspec2}) but now $k=\omega_n u_n$, where we
have introduced 
\be u_n = \sqrt{1-M^2/\omega_n^2} \en 
and $\omega_n=n\Omega$. The current and its Fourier transform remain
the same as in the massless case, and thus we obtain a very similar
formula \be
P_n = 2\frac{q^2\Omega^2}{\pi}n^2 \int_0^{u_n} du (\frac{1}{u^2}-1)
\J_n^2 (nuv_0), \label{harmasspow}\en
where only the upper bound in the integral has changed. The massless
gauge boson radiation (\ref{harmasspow}) is obviously recovered in the
limit $M\to 0$ or $n\to \infty$. Using the asymptotic expression
of the Bessel function (\ref{bessexp}), valid for $n\gg 1$, we obtain \be
P_n\approx \frac{q^2}{\pi^2}\Omega^2n \int_0^{u_n} \frac{du}{\sqrt{1-
u^2v_0^2}}\left(\frac{1-u^2}{u^2}\right) \exp\left(-\frac{2}{3}n(1-u^2v
_0^2)^{3/2}\right) \label{powexp} \en
In the case of massive gauge bosons with $M\gg a_m$, where $a_m$ is
the maximal acceleration, both functions in front of the exponent can
be taken at value $u=u_n$ and after integration we obtain \be
P_n \approx \frac{q^2}{2\pi^2}\frac{\gamma_0^2 M^2}{v_0^2} \left( 1-
\frac{M^2}{n^2\Omega^2}\right)^{-3/2} \left(1+\frac{\gamma_0^2v_0^2M
^2}{n^2\Omega^2}\right)^{-1} \frac{1}{n^2} \exp \left( -f(n)\right), 
\label{powmass} \en
where \be
f(n)= \frac{2}{3}\frac{n}{\gamma_0^3}\left(1+\frac{\gamma_0^2v_0^2M
^2}{n^2\Omega^2}\right)^{3/2}. \label{eq:f} \en
The function (\ref{eq:f}) has a minimum \be
f_m=\sqrt{3}\frac{Mv_0}{\gamma_0^2\Omega}  \label{eq:fmin} \en
which is reached at \be
n=n_m=\sqrt{2}\frac{\gamma_0v_0M}{\Omega}. \label{eq:nmax} \en
The second derivative of $f(n)$ at $n=n_m$ is \be
f_{nn}=\frac{2}{\sqrt{3}}\frac{\Omega}{\gamma_0^4v_0M}.\label{eq:fnn}\en
The spectrum (\ref{powmass}) has a maximum at $n \approx n_m$.

At very large $n \gg (M/a_m)n_m$ we obtain the spectrum (37) of a
massless case\footnote{This can not be seen directly from
(\protect\ref{powmass}) as the dominant term in this regime was
already neglected. To get this extra term, the factor $(1-u^2)$ in
(\protect\ref{powexp}) must be kept and integrated with the
exponential term}, as it must be according to (\ref{harmasspow}).

At $n_m \ll n<(M/a_m)n_m$ the energy flux is exponentially suppressed 
as \be
P_n \approx \frac{q^2}{2\pi^2}\frac{\gamma_0^2M^2}{v_0^2}\frac{1}{n^2}
\exp(-\frac{2}{3}n/\gamma_0^3). \label{eq:powlarge} \en

At $M/\Omega \ll n \ll n_m$ the energy flux is also exponentially 
suppressed as \be
P_n\approx \frac{q^2}{2\pi^2}\frac{\Omega^2}{v_0^4}\exp(-\frac{2}{3}M^
3v_0^3/n^2\Omega^3).\label{eq:powsmall} \en

In the region of the maximum $n_m$, the spectrum is \be
P_n\approx \frac{q^2}{12\pi^2}\frac{\Omega^2}{v_0^4}
\exp\left(-\sqrt{3}\frac{v_0M}{\gamma_0^2\Omega}\right)
\exp\left(-\frac{1}{2}f_{nn}(n-n_m)^2\right). \label{powmax} \en
Therefore, in the case we are considering $M\gg a_m$,  the 
characteristic energy of emitted gauge bosons is \be
E\approx n_m\Omega \approx \gamma_0M,  \label{eq:energy} \en
while in the massless case it is $E\approx \gamma_0a_m$.

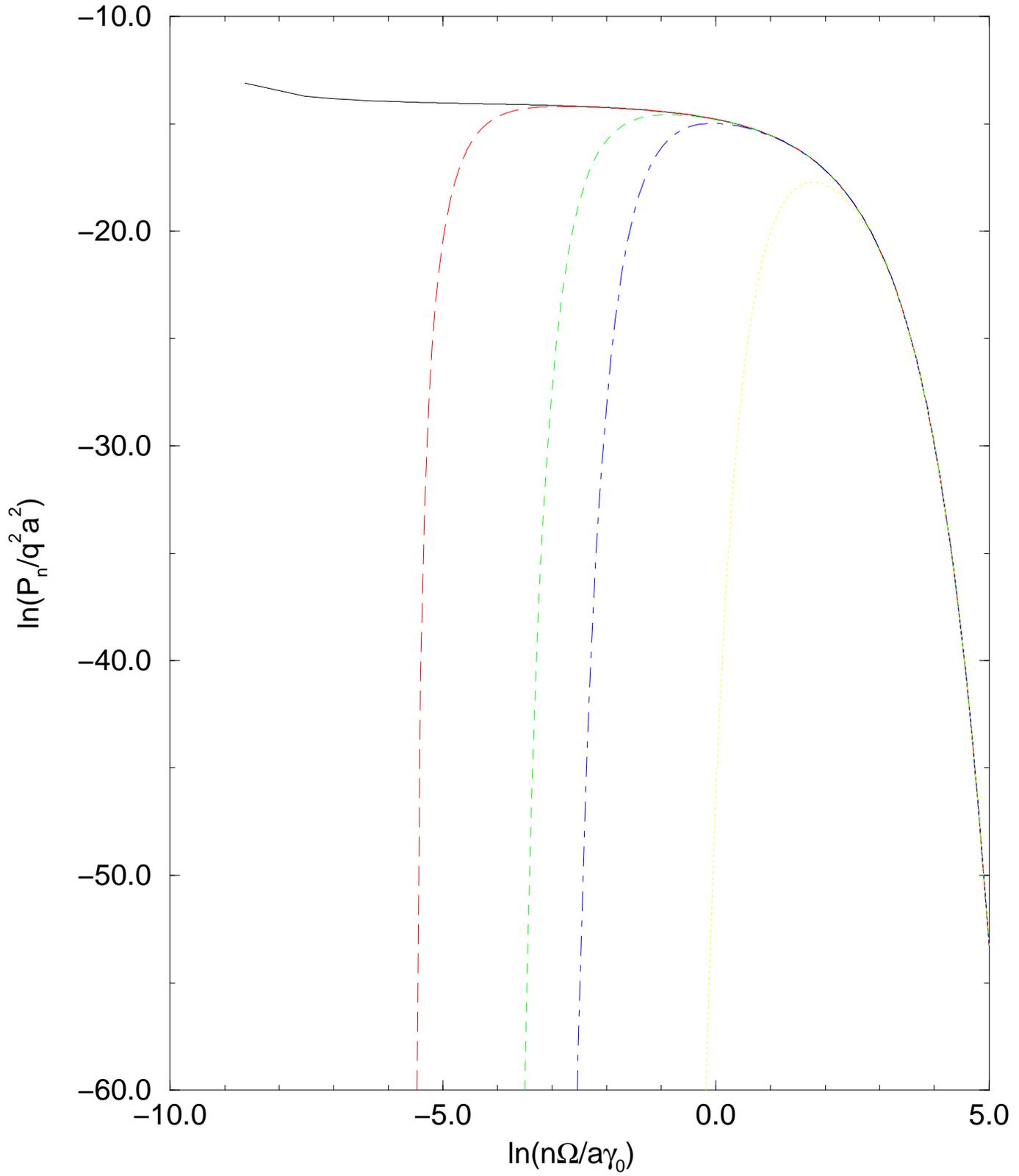
\begin{figure}
\caption{A log-log plot of the massive gauge quanta radiation spectrum
$P_n /(qa)^2$ for $\gamma_0 =25$ and $M/a_m$ taking the values $0.1$
(long dash), $0.5$ (dash), $1$ (dash-dot) and $5$ (dot). The massless
case from figure (\protect\ref{harmpow25}) has also been added (solid
line).}
\label{harmpow25M}
\end{figure}

The power spectra (\ref{harmasspow}) have been computed numerically
and are shown in Figure \ref{harmpow25M} for various values of
$M/a_m$.  Note that our analytical formulae given above are valid only
for $M/a_m \gg 1$. One can see that at large values of n
these spectra tend to the massless spectrum shown by a solid line.

The total radiated power, obtained by integrating (\ref{powmax}) \be
P\approx \sqrt{\frac{2\pi}{3}}\frac{q^2}{6\pi^2}
\frac{\gamma_0^3\Omega^2}{v_0^4} \sqrt{Mv_0/a_m}
\exp\left(-\frac{2}{3}\frac{Mv_0}{a_m}\right), \label{eq:powtot} \en
is exponentially suppressed in the case $M\gg a_m$.

The formulae above are valid for harmonically oscillating monopoles
with a 
frequency $\Omega$. We can reformulate these results for the instantaneous 
radiation power from a monopole moving with a proper accelleration $a$
and Lorentz 
factor $\gamma$. This is possible to do because for $M \gg a_m$ the 
radiation is strongly dominated by four very short bursts during the 
period $T$.   
The time dependence of the radiated power is determined mainly by the
factor 
$$
P(t) \propto \exp \left( -\frac{2}{3}\frac{Mv}{a(t)} \right),
$$
and the bursts occur at the maxima of the proper acceleration $a(t)$.
Expanding $a(t)$ near the maximum,
$a(t) \approx a_m+ \frac{1}{2}\ddot{a}_m(t-t_m)^2$, and using
Eq.(\ref{accprop}) we obtain an estimate for the 
duration of the bursts \be
\tau\sim\left( \frac{a_m}{|\ddot{a}_m|}\frac{3a_m}{Mv_0}\right) ^{1/2} \sim
\frac{1}{v_0\Omega \gamma_0}\left( \frac{a_m}{Mv_0} \right) ^{1/2}.
\label{dur} \en
It is $(a_m/Mv_0)^{1/2}$ times shorter than the corresponding 
duration in the massless case (\ref{deltat}).

The fact that the dominant part of the radiation comes out in short
bursts at $a \approx a_m$ suggests the following {\it ansatz} for the
instantaneous power:
\be P(a_m,\gamma,E)\approx 
k \frac{T}{4\tau} P_{har}(E,\Omega). \label{powa} \en
Here, $\gamma= \sqrt{\frac{2}{3}}\gamma_m$ is the Lorentz factor of a
harmonically oscillating monopole at the moment when it reaches
$a=a_m$, $k\sim 1$ is a numerical coefficient, and $P_{har}(E,\Omega)$
is given by 
(\ref{powmax}).  We expect Eq.(\ref{powa}) to be valid for $E$ near
the maximum of the spectrum at least for $|E-E_m|/E_m \ll 1$. 
Integrating the instantaneous spectrum (\ref{powa}) over the
period $T$, with $a(t)$ and $\gamma(t)$ for the harmonic oscillator
motion, and comparing with the averaged oscillator spectrum
(\ref{powmax}) we determine the
normalizing 
constant, $k=2\sqrt{2}/(3\sqrt{\pi})$. The total 
power and the spectrum for a 
relativistic ($\gamma \gg 1$) monopole moving with proper acceleration
$a$ and Lorentz-factor $\gamma$ are thus given by
\be P(a)=\frac{\sqrt{3}}{8\pi}q^2aM\exp\left( - \frac{2}{3}\frac{M}{a}
\right), \label{powt} \en
\be P(a,\gamma,E)=\frac{q^2}{12\pi\sqrt{\pi}}\frac{a}{\gamma}
\left( \frac{M}{a} \right)^{1/2}\exp\left( -\frac{2}{3}\frac{M}{a}\right)
\exp \left( -f_{EE}(E-E_m)^2\right), \label{powE} \en
where 
\be E_m=\sqrt{3}\gamma M. \label{Emax} \en
\be f_{EE}= \frac{4}{27} \frac{1}{Ma\gamma^2}, \label{width} \en
Eqs.(\ref{Emax}) and (\ref{width}) indicate that the gauge quanta
are emitted with a characteristic energy $E\sim \gamma M$ in a narrow
range
\be
\Delta E/E\sim (a/M)^{1/2}.
\en

In the case of a vector field $A_{\mu}(x)$ considered above, there is a
compensation 
in the energy-momentum tensor between the two terms corresponding to two 
nonvanishing components of the field  (e.g $A_0$ and $A_3$). 
Such a compensation is absent 
for a scalar field $\phi(x)$ with an interaction $L(x)=q\phi(x)j(x)$,
where
\be
j(x)=\int d\tau\delta^{(4)}(x-x(\tau)) \en
and $\tau$ is the proper time along the trajectory of the source.
In this case, a similar analysis, for $M\gg a$,  gives the following
expression for the instantaneous 
radiation power of a monopole moving with acceleration $a$ and Lorentz-factor
$\gamma \gg 1$: \be
P_s = \frac{3\sqrt{3}}{8\pi}q^2\gamma^2a M
\exp\left(-\frac{2}{3}\frac{M}{a}\right).\en

\section{Production of high energy hadrons.} \label{sec5}

We shall consider here the production of high energy hadrons by an
accelerating monopole through the most relevant process for this
purpose: emission of "heavy" off-mass-shall gluons.

The GUT monopole has \cite{Preskill} a chromomagnetic
charge, which is screened beyond the confinement radius. The
monopole-gluon coupling constant is $1/\alpha_{QCD}$. Production of
off-mass-shall gluons occurs in the same way as the radiation of
massive gauge bosons (Section IV).  The virtualities of gluons $Q^2$
higher than $a^2$ are exponentially suppressed and thus the typical
vituality and energy of the emitted gluon is $a^2$ and $\gamma a$,
respectively. A virtual gluon decays into two partons, thus starting a
parton cascade, which is described by the
Gribov-Lipatov-Alterelli-Parisi equation \cite{Gribov}.  Beyond the
confinement radius, this cascade is converted into a hadronic jet.
This picture is quite similar to hadron production in $e^+e^-$
annihilation or in deep-inealastic scattering.  We expect it to apply
as long as $a>\Lambda$, where $\Lambda=0.234~GeV$ is the QCD scale.

The spectrum of hadrons $h$ radiated by a monopole moving with a
proper acceleration $a$ can be expressed with the help of a
fragmentation function, $W_h(a,E_g,x)$ as \be
N_h(E_h)=\int_{E_h}^{E_g^{max}} dE_g N_g(E_g)W_h(a,E_g,x)/E_g, \label{hadspec} \en
where $x=E_h/E_g$ and $N_g(E_g)$ is the spectrum of heavy gauge bosons 
calculated in Section V. Actually one can use the massless spectrum when 
considering virtualities less than $a^2$. In the rest frame of the monopole,  
the typical virtuality and energy of the gluon are, respectively, $a^2$
and $a$, similar to the virtual photon in $e^+e^-$-annihilation. We
shall use the fragmentation function based on the calculations of
ref.\cite{Webber} (see also more general expression in
ref.\cite{Doksh}) which depends only on the parameter $x$: \be
W_h(x)=\frac{K_h}{x}\exp\left(-\frac{\ln^2x/x_0}{\sigma_0}\right),
\label{frag} \en 
where $K_h$ is the normalization constant, taken from the condition \be
\int_0^1 x W_h(x) dx=f_h, \en
where $f_h$ is the fraction of energy transferred to hadrons of type $h$,
$x_0=\sqrt{\Lambda/a}$, and 
\be \sigma_0=\frac{1}{12 b}\sqrt{\frac{2\pi}{3\alpha_s^3}}. \en
The value $b=(1/12\pi)(33-2f)$ determines the QCD coupling constant in
the one-loop approximation as $\alpha_s^{-1}=b\ln(a^2/ \Lambda^2)$,
where $f$ is the number of guark flavors unfrozen at a given value of
$a$. Numerically, $\sigma_0=0.121 \sqrt{b}\ln^{3/2}(a^2/\Lambda^2)$.
Note that the fragmentation function (\ref{frag}) differs
significantly from the function introduced by Hill \cite{H+} which was
subsequently used in much of the literature on cosmic rays from
topological defects.

The fragmentation function $W_h(x)$ has a maximum at \be
x=x_m=x_0\exp(-\sigma_0/2). \en
The origin of this maximum is related to the effect of coherent
radiation of soft gluons by a jet of partons
\cite{Webber},\cite{Doksh}.  At small $x$ the size of emitted gluons
is larger than the width of the parton beam. A jet therefore radiates
soft gluons as a single source with a color charge equal to the
algebraic sum of the color charges of partons in the beam.

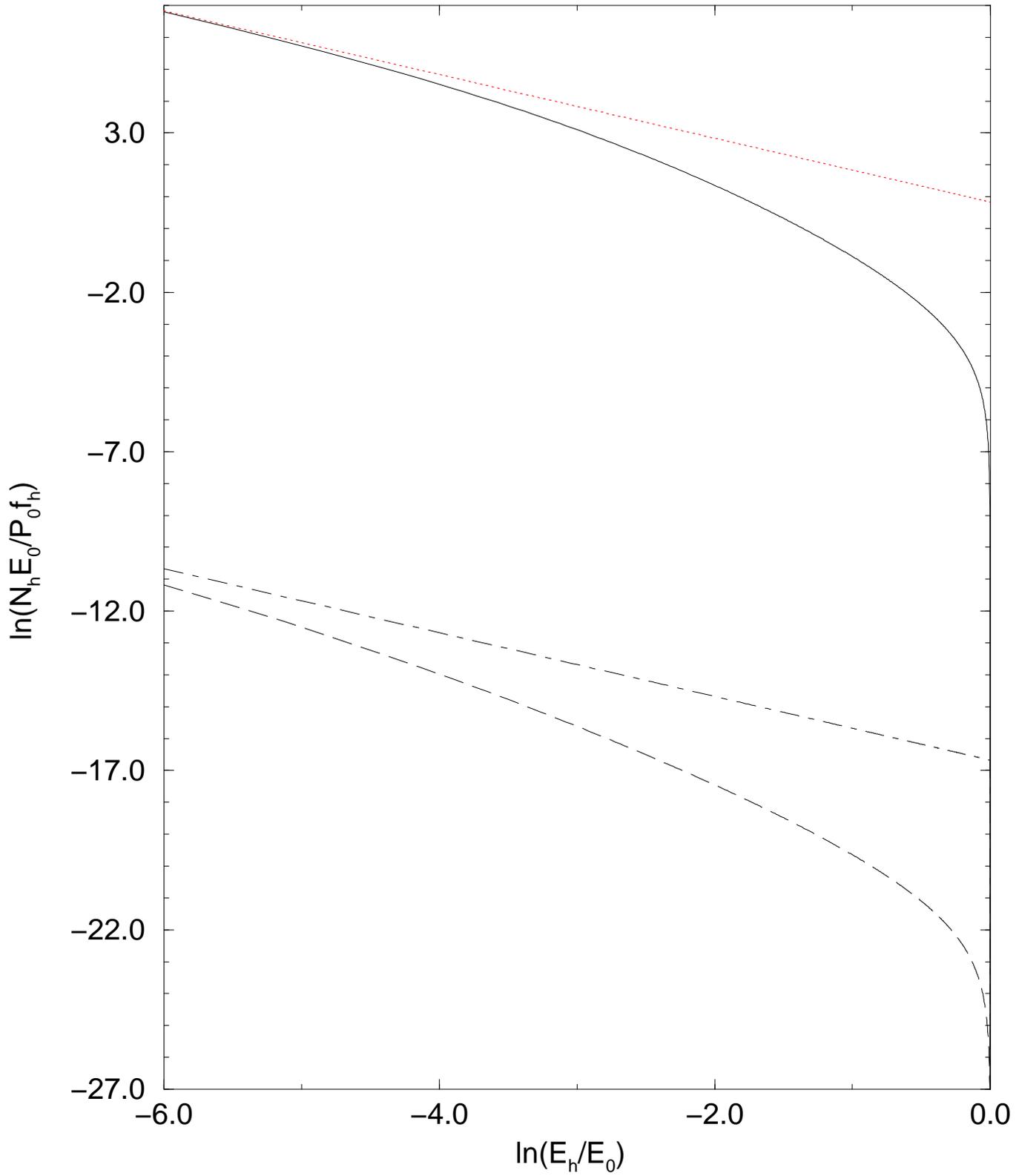
\begin{figure}
\caption{A log-log plot of the hadron spectrum
(\protect\ref{hadrspec}) for $a=100~GeV$ (solid line) and $a=10~TeV$
(dashed line) with their low energy approximations
(\protect\ref{lowhadr}) (respectively dotted line, dash-dotted line).}
\label{hadrspec10}
\end{figure}

For ultrarelativistic monopoles, the approximation (\ref{specap}) for
the spectrum of massless gauge bosons can be used instead of the exact
spectrum. Then, the integration in Eq. (\ref{hadspec}) for the hadron
spectrum can be performed and we obtain, for $E_h \leq E_0$, \be 
N_h (E_h)= \frac{P_0}{E_0} f_h \frac{E_0}{E_h} \frac{\e^{-\sigma_0^2
/4}}{x_0} \frac{\phi (\ln (x_0)/\sqrt{\sigma_0})-\phi (\ln (E_0 x_0
/E_h)/\sqrt{\sigma_0})}{\phi (\sqrt{\sigma_0}/2)}, \label{hadrspec} \en 
where we have introduced the auxiliary function \bea
\phi (x) & = & \int_x^{+\infty} \e^{-t^2}dt .\ena
We obviously have $N_h(E_0)=0$, while the low energy asymptotic
behavior of the spectrum is: \be
N_h(E) \approx \frac{P_0}{E_0} f_h\frac{\e^{-\sigma_0^2
/4}}{x_0} \frac{\phi (\ln (x_0)/\sqrt{\sigma_0})}{\phi (\sqrt{\sigma
_0}/2)}  \frac{E_0}{E_h}. \label{lowhadr} \en
A numerical example for the hadron (nucleon) spectrum, which
corresponds to $a=100~GeV$ and $a=10~TeV$ is shown in figure
\ref{hadrspec10}. The value of $f_h$ can be taken as 0.1, which is
roughly valid for nucleons. The shape of the spectrum depends only on
the parameter $a$.

\section{Concluding remarks} \label{sec6}

We have analyzed the radiation of massless and massive gauge bosons by
accelerating monopoles.  In the massless case, we calculated the
radiation spectrum for an oscillating $M{\bar M}$ pair connected by a
straight string and for a monopole undergoing a harmonic oscillator
motion.  Combining this with the standard analysis of the synchrotron
radiation, we arrive at the following qualitative picture.

For a monopole of magnetic charge $q$ moving with a proper
acceleration $a$, the total radiated power is $P=q^2a^2/6\pi$.
In our case, $a=\mu/m$, where $\mu$ is the string tension and $m$ is
the monopole mass.
In the instantaneous rest frame of the monopole, the characteristic 
energy of the emitted quanta is ${\bar E}_{rest}\sim a$, and thus in the
observer's frame it is 
\be
{\bar E}\sim\gamma a, 
\label{bare}
\en
where $\gamma$ is the Lorentz
factor of the monopole.  For $E\gg{\bar E}$ the spectral power is strongly
suppressed, while for $E<{\bar E}$ the spectrum is somewhat dependent
on the type of motion (for example, the spectrum is flat for a
harmonic oscillator and has a maximum at $E\sim{\bar E}$ for a
circular motion).

The same picture applies in the case of massive gauge bosons, as long
as their mass is sufficiently small, $M\ll a$.  In the opposite limit
of large mass, $M\gg a$, the radiation power is exponentially
suppressed [see Eq.(\ref{eq:powtot})], and gauge quanta are emitted
with a characteristic energy ${\bar E}\sim\gamma M$ in a narrow range
$\Delta E/E\sim (a/M)^{1/2}$. The interesting case for practical
applications is the production of hadrons by an accelerating
monopole. It effectively occurs through radiation of gluons due to the
chromodynamic charge of the monopole. As in the general case of heavy
quanta, the radiation of gluons with virtualities $|Q|^2 >a$ is
exponentially suppressed and therefore most gluons are produced with
$|Q|^2 \leq a$, where the massless limit is approximately
valid. Beyond the confinement radius, high-energy gluons fragment into
hadrons; the corresponding fragmentation function is given by
Eq.(\ref{frag}).

The obtained formulae for radiation of accelerating monopole can be 
straightforwadly generalized to any other accelerating particle.

In a monopole-string network, with $N$ strings attached to each
monopole, the proper acceleration of a monopole is determined by
the vector sum of the tension forces exerted by the strings.  By order
of magnitude it is still given by $a\sim\mu/m$.  The cosmological
evolution of monopole-string networks is expected to be characterized
by a single length scale, 
\be
\xi(t)\sim\kappa t, 
\label{xi}
\en
where $\kappa=const$
and $t$ is the cosmic time.  This scale gives the average distance
between monopoles and the average length of string segments.  The
typical energy of a monopole is
\be
E_m\sim\mu\xi\sim\mu\kappa t.
\label{emt}
\en
The corresponding Lorentz factor is $\gamma \sim (\mu/m)\kappa t$, and
from Eq.(\ref{bare}) the typical energy of quanta emitted at time $t$ is
\be
{\bar E}\sim (\mu^2/m^2)\kappa t.
\label{et}
\en

Assuming that radiation of gauge quanta is the dominant energy loss
mechanism of the networks, one finds \cite{vv} that the parameter
$\kappa$ is given by
\be
\kappa\sim\mu/\alpha m^2,
\label{kappa}
\en
where $\alpha=e^2/4\pi$ and $e$ is the gauge coupling.  Networks can
also lose energy by production of closed loops of string and of small
nets.  The effect of these mechanisms is hard to estimate without
numerical simulations of the network dynamics.  For the time being,
Eq.(\ref{kappa}) should be regarded as a lower bound on $\kappa$, the
upper bound being due to causality, $\xi \lesssim t$.  Hence,
\be
\mu/\alpha m^2 \lesssim \kappa \lesssim 1.
\label{kappa'}
\en

It is clear from Eqs.(\ref{emt}),((\ref{et}) that the energies of
monopoles themselves and of the quanta they emit can get arbitrarily
large at sufficiently late times.  In particular, the emitted
particles can have super-Planckian energies, $E\gg m_p$, where $m_p$
is the Planck mass.  To give a quantitative estimate, we note that the
string tension $\mu$ and the monopole mass $m$ can be expressed in
terms of the corresponding symmetry breaking scales as
$\mu\sim\eta_s^2$, $m\sim 4\pi\eta_m/e$.  These scales are expected to
satisfy 
\be
\eta_{EW} \lesssim\eta_s<\eta_m<m_p,
\en
where $\eta_{EW}\sim 10^2 GeV$ is the electroweak symmetry breaking
scale.  It follows from Eqs.(\ref{et}),(\ref{kappa'}) that at the
present time
\be
{\bar E}/m_p \gtrsim 10^{56}(\eta_s^6/\eta_m^4m_p^2),
\en
and thus the emitted particles are super-Planckian provided that
\be
\eta_s^3/\eta_m^2 m_p\gg 10^{-28}.
\en
This covers a wide range of parameter values.  For example, one could
have GUT-scale monopoles, $\eta_m\sim 10^{16}~GeV$, and
intermediate-scale strings, $\eta_s \gtrsim 10^9~GeV$, or
electroweak-scale strings and light monopoles with $\eta_m \lesssim
10^7~GeV$.  

The interaction of super-Planckian particles with the microwave
background and with cosmic magnetic fields, the resulting cosmic
ray fluxes, and some other defect models that can give rise to such
particles will be discussed elsewhere \cite{bv}.  Conventional
astrophysical acceleration mechanisms are not capable to account for
particles of such tremendous energy, and if they are ever observed, it
appears that topological defects would be the only possible
explanation. 

\section*{Acknowledgements}
One of the authors (V.B.) is grateful to B.R.Webber for providing the 
manuscript of the book [15] before the publication  and to B.R.Webber and 
V.A.Khose for useful discussions. The two other authors (A.V. and X.M.)
were supported by the National Science Foundation.

\appendix
\section{Gauge boson radiation spectrum} \label{bospow}
The electromagnetic radiation of a periodic electromagnetic source has
been extensively studied and is well understood \cite{ll}. However, for the
purpose of this work we will need a generalization of the standard
formulae to the case of radiation of {\it massive} gauge bosons.

We consider a gauge boson field $A^\mu$ of mass $M$
coupled to a source term $j^\mu$, \be
{\cal L} =-\frac{1}{2} F^{\mu\nu} F_{\mu\nu} +M^2 A^\mu A_\mu +A_\mu
j^\mu .\en
To this Lagrangian correspond the equations of motion \be
(\Box +M^2 )A^\mu =j^\mu ,\label{masseq} \en
and the energy momentum tensor for the bosons \be
T_{\mu\nu}=-2(\partial_\mu A_\lambda )(\partial_\nu A^\lambda )-{\cal
L} g^{\mu\nu}.\en
The power radiated in bosons can be found from
\be
P(t)=r^2 \int d\Omega S_r (t),\en
where \be 
S_r = -2(\partial_0 A_\lambda )(\partial_r A^\lambda ) \label{rflux}\en
is the radial outgoing energy flux and $\partial_r$ is a derivative
with respect to the radial distance $r=|{\bf x}|$. For a periodic motion,
this power can be averaged over a period to get \be
P=r^2 \int \frac{dt}{T} \int d\Omega S_r (t),\label{totpowmass} \en

In the case of a massless gauge bosons $M=0$, the equation
(\ref{masseq}) can be solved exactly using the Lienhart-Wiechert
potentials, which greatly simplifies the problem. For a massive gauge 
boson there is no such solution.  However, Eq. (\ref{masseq}) is
linear and can be solved in Fourier space. We therefore
introduce the Fourier transforms \bea
A^\mu (\omega_n ,{\bf k}) & = & \frac{1}{T} \int dt \int d^3 x \e^{i(\omega_n
t-{\bf k}\cdot {\bf x})} A^\mu (t,{\bf x}),\\
j^\mu (\omega_n ,{\bf k}) & = & \frac{1}{T} \int dt \int d^3 x \e^{i(\omega_n
t-{\bf k}\cdot {\bf x})} j^\mu (t,{\bf x}), \ena
where $T$ is the period of the source and $\omega_n =2\pi n/T$.
Eq. (\ref{masseq}) becomes \be
(k^2-\omega_n^2 +M^2)A^\mu (\omega_n {\bf k})=j^\mu (\omega_n,{\bf k}).\en
This equation is trivially solved to get \bea
A^\mu ({\bf x}) & = &\sum_n \e^{-i\omega_n t} A_n^\mu ({\bf x}) \\
A_n^\mu ({\bf x}) & = & \int \frac{d^3 k}{(2\pi )^3}\e^{i{\bf k}\cdot
{\bf x}} \frac{j^\mu (\omega_n,{\bf k})}{k^2-k_0^2\pm i \varepsilon},
\label{anx} \ena where we have introduced \be
k_0^2 = \omega_n^2 -M^2 .\en
The expression (\ref{anx}) for $A_n^\mu ({\bf x})$ can be greatly
simplified in the particular case we are interested in, where it is
evaluated at a point ${\bf x}$ far from the source so that at any
point of the source ${\bf x}'$, $|{\bf x}|\gg |{\bf x'}|$: \bea
A_\mu (\omega_n,{\bf x}) & = & \int \frac{dt}{T}\int d^3 x' j^\mu
(t,{\bf x'})\int \frac{k^2 dk}{k^2-k_0^2\pm i\varepsilon} \int \sin 
\theta d\theta d\phi \e^{ik|{\bf x}-{\bf x}'|\cos \theta} \\
& = & \int \frac{d^3x'\e^{ik_0|{\bf x}-{\bf x}'|}}{4\pi|{\bf x}-{\bf 
x}'|}j^\mu (t,{\bf x}') \approx \frac{\e^{ik_0r}}{4\pi r} j^\mu 
(\omega_n,k_0{\bf n}),\ena
where we have introduced $r=|{\bf x}|$ and ${\bf n}={\bf x}/r$.
The radial energy flux (\ref{rflux}) can then be expressed to first
order in $1/r$ as \be
S_r = -\frac{1}{8\pi^2 r^2} \sum_{n,m} (\omega_n k_{0}) \e^{i(m-n)
\omega_n t} j^\mu (\omega_n,k_{0} {\bf n}) j^*_\mu (\omega_m,k
_{0} {\bf n}).\en
Then the total power (\ref{totpowmass}) becomes \bea
P & = & \sum_n P_n ,\\
P_n & = & \frac{k_0\omega_n}{8\pi^2}j^\mu (\omega_n,k_{0} {\bf n})
j^*_\mu (\omega_n,k_{0} {\bf n}) \ena
where $P_n$ is obviously the power radiated by the mode $n$.

\newpage

\end{document}